# Automatic Evolution of Machine-Learning based Quantum Dynamics with Uncertainty Analysis

Kunni Lin[1,2], Jiawei Peng[1,2], Chao Xu[2,3], Feng Long Gu[2,3,*] and Zhenggang Lan[2,3,*]

*1 School of Chemistry, South China Normal University, Guangzhou 510006, P. R. China.*

*2 MOE Key Laboratory of Environmental Theoretical Chemistry, South China Normal University, Guangzhou 510006, P. R. China.*

*3 SCNU Environmental Research Institute, Guangdong Provincial Key Laboratory of Chemical Pollution and Environmental Safety, School of Environment, South China Normal University, Guangzhou 510006, P. R. China.*

*\* Corresponding Author. E-mail: gu@scnu.edu.cn; zhenggang.lan@m.scnu.edu.cn.*

**Abstract**

The machine learning approaches are applied in the dynamical simulation of open quantum systems. The long short-term memory recurrent neural network (LSTM-RNN) models are used to simulate the long-time quantum dynamics, which are built based on the key information of the short-time evolution. We employ various hyperparameter optimization methods, including the simulated annealing, Bayesian optimization with tree-structured parzen estimator and random search, to achieve the automatic construction and adjustment of the LSTM-RNN models. The implementation details of three hyperparameter optimization methods are examined, and among them the simulated annealing approach is strongly recommended due to its excellent performance. The uncertainties of the machine learning models are comprehensively analyzed by the combination of bootstrap sampling and Monte-Carlo dropout approaches, which give the prediction confidence of the LSTM-RNN models in the simulation of the open quantum dynamics. This work builds an effective machine learning approach to simulate the dynamics evolution of open quantum systems. In addition, the current study provides an efficient protocol to build the optimal neural networks and to estimate the trustiness of the machine learning models.

**I. Introduction**

With the rapid development of computer artificial intelligence, machine learning (ML) plays more and more important roles in theoretical chemistry, including the construction of the molecular Hamiltonian,[1-10] the analysis of trajectory-based molecular dynamics evolution,[11-15] the prediction of molecular properties and chemical reactions,[16-20] the design of novel functional materials.[21-22] In addition to these, considerable efforts were made to employ the ML approach to simulate dynamics evolutions recently.[23-35]

The quantum evolution of the nonadiabatic dynamics of high-dimensional systems is always a key research object in theoretical simulations,[36-39] due to their important roles in chemistry, physics and biology. The system-plus-bath model is widely used to treat the dynamics of complex systems,[36] in which the reaction center is treated as the reduced system and its dynamics is studied in detail, while a large number of surrounding degrees of freedom are treated as environment. The theoretical description of the dissipative dynamics of open quantum systems has been received considerable attentions over several decades.[37-58] Although many theoretical approaches, ranging from full quantum[37, 43-50, 59] to semiclassical and mixed-quantum-classical dynamics approaches,[38-39, 51-54, 58, 60] were developed to simulate the reduced dynamics of open systems, all of them may perform well in some situations but suffer from some deficiencies in other cases.[51, 56-57, 61-62] For instance, numerically exact dynamics approaches, such as the hierarchical equations of motion (HEOM)[37, 40-42] and the hybrid stochastic-deterministic HEOM,[63] the multiconfigurational time-dependent Hartree (MCTDH),[43-46, 59] tensor-network decomposition,[47-49, 64-67] and so on,[50] may give the correct description of the open quantum dynamics, while the employment of them to solve realistic problems often suffers from either numerical convergence problems or high computational costs.

In these years, several theoretical efforts tried to employ the ML approaches to simulate the quantum dynamics of reduced systems.[23-34] Some works tried to build a ML model based on the short-time dynamics and then use it to predict the long-time evolution. These treatments are highly correlated to the transfer tensor method originally proposed by Gerrillo and Cao.[68] In the transfer tensor approach, a linear dynamics map was built from the historical short-time dynamics. When all critical dynamical information is compressed into the transfer tensor, the long-time reduced dynamics can effectively be simulated by such transfer tensor under the assumption of the time-translational invariance. This interesting idea aroused the further research attentions. For example, it is possible to combine the transfer tensor method with the mixed quantum-classical Liouville dynamics, as shown by Geva, Cao and co-workers.[69] The transfer tensor formulism is a linear map model, while it is possible to build a nonlinear dynamical map model. Along this idea, different ML approaches, such as recurrent neural networks (RNN),[25-26, 28, 32-33, 70-71] convolutional neural networks (CNN)[28, 30] and kernel ridge regression (KRR)[27] were used to build such nonlinear map models. For instance, Zhao and co-workers[25-26] took the RNN[70] and long short-term memory recurrent neural networks (LSTM-RNN)[72] in the simulation of the evolution of open quantum systems. And they also combined the CNN and the fully-connected neural networks (FCNN) to predict the non-adiabatic dynamics of a paradigmatic model with the Landau-Zenner transition.[29] Lin et. al[23] tried to access the confidence interval of the forecasting long-term dynamics in the LSTM-RNN simulation by addressing the prediction uncertainty with the bootstrap resampling technology.[73-76] Wu et. al[28] proposed the hybrid CNN-LSTM framework to achieve the higher prediction accuracy of the long-term quantum dynamics. In similar but different approaches by Rodríguez[30] and Ullah,[27] a group of system-plus-bath models with different parameters were taken and their exact dynamics results are collected. After all dynamical evolutions were broken into many small pieces of time durations, a unified ML model (CNN by Rodríguez[30] and KRR by Ullah[27]) was built to simulate the long-term quantum evolution of other similar reduced models, when their early-time dynamics are known. Ullah et. al[77] employed the CNN model in the simulation of the exciton dynamics in the light-harvesting complexes and confirmed the excellent performance of this artificial-intelligence based open quantum dynamics approach. Banchi et. al[24] proposed to employ the RNNs with Gated Recurrent Unit (GRU)[78] to simulate the non-Markovian quantum processes, starting from various initial conditions. Secor et. al[79] recently demonstrated that the quantum propagator can be represented by the FCNN. Choi et. al showed that the unsupervised ML models can be used to represent the major features in quantum evolutions of qubit systems.[71]

No matter which ML models were used to treat the quantum dynamics of open quantum systems, some practical implement problems must be concerned. One major question is how to set up the proper neural network (NN) structures in the training process. For examples, how many layers should be chosen and how many neurons should be given for each layer. These parameters named as the "hyperparameters" determine the accuracy of the NN models.[80-81] Normally, they are given before the training step, and the real parameters (such as weight or bias at each neuron) are determined by the regression. Only when several regression processes with different hyperparameters are performed, the suitable hyperparameter set can be determined according to the model performance. It is not trivial to obtain the reasonable hyperparameter set, particularly when many hyperparameters must be adjusted in the deep learning. Thus, the automatic and efficient hyperparameter optimization approaches are highly useful. The second critical issue in the application of the NN model is relevant to the proper definition of the model uncertainty. Take the prediction of the dynamical evolution as an example. When the further dynamics is already known, it is easy to use the long-time dynamics data to benchmark the prediction results of ML models. However, in reality such long-time dynamical evolution is completely missing before the ML forecast, and we have no data of future dynamics to clarify the reliability of the ML models. In such case, if the model uncertainty can be derived formally in mathematics view, we may get the primary assessment on the confidence of the ML model. This gives us the important rule to judge the reliability of the ML model in long-time dynamics simulation of open quantum system. Therefore, we must find the proper way to estimate the model uncertainty.

In the current work, we tried to address the above two essential issues in details. We tried to employ the LSTM-RNN approach[70, 72] to learn the short-time quantum dynamics evolution and used the trained ML models to predict the long-time quantum evolution. The exact quantum dynamics was obtained by using the time-dependent tensor-train approach.[47-49, 64-67]

In the construction of the LSTM-RNN models, we tried to use three automatic approaches to optimize the hyperparameters, including simulated annealing (SA),[82-84] Bayesian optimization with tree-structured parzen estimator (BO-TPE)[85-87] and random search (RS) methods.[87-89] These methods allow us to directly obtain the optimal hyperparameters including the NN topology. These three approaches were widely used in the ML field[85, 88, 90-94] and some of them were already applied in quantum chemistry. For instance, Westermayr et. al [95] employed the RS to optimize the hyperparameters in the construction of the deep learning NNs that were used to simulate the nonadiabatic excited-state molecular dynamics. Deng et. al[96] proposed that the Bayesian optimization is an effective approach in the solution of inverse problems in time-dependent quantum dynamics. Except them, other algorithms are also available.[97-98] To address the ML model uncertainty, two popular Monte-Carlo (MC) approaches, namely bootstrap resampling[73-76] and MC dropout,[99-102] were chosen to evaluate the confidence interval of model prediction. Similar approaches were occasionally employed in theoretical chemistry community. For example, Peterson et. al[99] implemented the bootstrap resampling approach in the uncertainty analysis of the NN-based potential energy surfaces. Wang et. al[103] used the deep NN model with the MC dropout approach to analyze the power degradation tendency in the proton-exchange-membrane fuel cell, along with the prediction interval.

The present work provides us a practical step-by-step protocol, which allows us to build the reasonable ML models in an automatic manner, to propagate the quantum dynamics evolution efficiently and to access its uncertainty reliably. This paves a rather clear way to implement the ML model to treat the quantum evolution. In addition, we provide very detailed discussions on how to perform the automatic hyperparameters optimization and evaluate the ML model uncertainty. Since these two issues must be considered in realistic applications, the current work is very helpful to the future research works that use supervised ML models to study other physical-chemical problems.

**II. Methods**

*2.1. The Hamiltonian.*

In current work, we considered a system-plus-bath model, in which the system part is composed of two local-excited (LE) electronic states and the bath part includes many vibrational modes, i.e.

$$H = H_S + H_B + H_{SB}. \tag{1}$$

The system part is written as

$$H_S = \sum_{k=1}^{2} |\varphi_k\rangle V_{kk} \langle\varphi_k| + \sum_{k \neq l} |\varphi_k\rangle V_{kl} \langle\varphi_k|, \tag{2}$$

where $V_{kk}$ represents the energy of the excited state $|\varphi_k\rangle$ and $V_{kl}$ represents the electronic coupling term. Here we assume that two LE electronic states couple with their own individual bath modes. The bath Hamiltonian is defined as

$$H_B = \sum_{k=1}^{2} \sum_{j}^{N_b} \frac{1}{2} \omega_{kj} (Q_{kj}^2 + P_{kj}^2). \tag{3}$$

The bilinear system-bath interaction is considered as

$$H_{SB} = \sum_{k=1}^{2} |\varphi_k\rangle (\sum_{j}^{N_b} \kappa_{kj} Q_{kj}) \langle\varphi_k|. \tag{4}$$

In Eq.3 and 4, $N_b$ represents the total number of bath modes coupled with the single LE state. Three parameters, $\omega_{kj}$, $Q_{kj}$, and $P_{kj}$, are the corresponding frequency, position and momentum of each bath mode, respectively. The $\kappa_{kj}$ characterizes electron-phonon coupling strength. The subscripts $k$ and $j$ refer to the indices of the electronic state and the bath mode, respectively. The bath is characterized by the Debye-type spectral density:

$$J(\omega) = \frac{2\lambda\omega\omega_c}{\omega^2 + \omega_c^2}, \tag{5}$$

Where $\omega_c$ and $\lambda$ refer to the characteristic frequency and the reorganization energy, respectively. The spectral density is represented by a series of discretized bath modes:

$$J_k(\omega) = \frac{1}{2}\pi \sum_{i=1}^{N} \kappa_{ki}^2 \delta(\omega - \omega_{ki}). \tag{6}$$

The coupling strength of each mode $\kappa_{ki}$ is evaluated by the following equation:

$$\kappa_{ki} = (\frac{2}{\pi} J_k(\omega_{ki}) \Delta \omega)^{1/2}, \tag{7}$$

when the sampling interval $\Delta \omega$ is given.

The electronic ground state is described by a multi-dimensional harmonic potential with the minimum at $Q_{kj} = 0$. The initial condition in dynamics is defined as the vertical excitation of the lowest vibrational level of the electronic ground state to one LE state.

In current work, we considered symmetric and asymmetric site-exciton models with different parameters, and all detailed information are given in Table S1 in Supporting Information (SI).

*2.2. Tensor-Train.*

In the tensor-train framework, a quantum state is reformulated as a matrix-product state (MPS)[49]:

$$|\Psi\rangle = \sum_{\{s_i\}} \mathbf{A}^{s_1} \cdots \mathbf{A}^{s_i} \cdots \mathbf{A}^{s_m} |s_1 \cdots s_i \cdots s_m\rangle, \tag{8}$$

where $\{|s_i\rangle\}$ represents $m$ local basis. $\mathbf{A}^{s_i}$ is a site-dependent rank-3 tensor with dimensions $\alpha_{i-1} \times s_i \times \alpha_i$ ($\alpha_0 = \alpha_m = 1$). Based on the gauge transformation, the above

MPS can be reconstructed as multiform, particularly, the left (right) canonical MPS in terms of the left (right) orthonormal tensor component $\mathbf{L}^{s_i}$ ($\mathbf{R}^{s_i}$), or the mixed-canonical MPS:

$$\begin{aligned} |\Psi\rangle &= \sum_{\{s_i\}} \mathbf{L}^{s_1} \cdots \mathbf{L}^{s_i} \cdots \mathbf{L}^{s_m} |s_1 \cdots s_i \cdots s_m\rangle \\ &= \sum_{\{s_i\}} \mathbf{R}^{s_1} \cdots \mathbf{R}^{s_i} \cdots \mathbf{R}^{s_m} |s_1 \cdots s_i \cdots s_m\rangle \\ &= \sum_{\{s_i\}} \mathbf{L}^{s_1} \cdots \mathbf{L}^{s_{i-1}} \mathbf{M}^{s_i} \mathbf{R}^{s_{m-1}} \cdots \mathbf{R}^{s_m} |s_1 \cdots s_i \cdots s_m\rangle, \\ &= \sum_{\{\alpha_{i-1}, s_i, \alpha_i\}} \left[\mathbf{M}^{s_i}\right]_{\alpha_{i-1}, \alpha_i} \left|\Psi_{L,\alpha_{i-1}}^{[1:i-1]}\right\rangle |s_i\rangle \left|\Psi_{R,\alpha_i}^{[i+1:m]}\right\rangle \end{aligned} \quad (9)$$

where $\mathbf{M}^{s_i}$ represents the corresponding tensor on the active site $i$; $\left|\Psi_{L,\alpha_{i-1}}^{[1:i-1]}\right\rangle$ and $\left|\Psi_{R,\alpha_i}^{[i+1:m]}\right\rangle$ are effective states generated from the left and right orthonormal basis, respectively.

Similarly, an operator is also redefined in terms of the matrix product operator (MPO) within $m$ local basis:

$$\hat{O} = \sum_{\{s_i\},\{s_i'\},\{\beta_i\}} \mathbf{W}_{\beta_0,\beta_1}^{s_1,s_1'} \cdots \mathbf{W}_{\beta_{i-1},\beta_i}^{s_i,s_i'} \cdots \mathbf{W}_{\beta_{m-1},\beta_m}^{s_m,s_m'} |s_1 \cdots s_i \cdots s_m\rangle \langle s_1' \cdots s_i' \cdots s_m'|, \quad (10)$$

where $\mathbf{W}^{s_i,s_i'}$ is a site-dependent rank-4 tensor with dimensions $\beta_{i-1} \times s_i \times s_i' \times \beta_i$ ($\beta_0 = \beta_m = 1$). Several avenues are proposed to construct the MPO.[49, 104] In this work, the site-exciton model Hamiltonian was rewritten with the second quantization, as the occupation number representation is convenient for the MPO construction.[64-67, 105]

Based on the defined MPS and MPO, several algorithms were developed to solve the time-dependent Schrödinger equation. In this work, the time-dependent variational principle (TDVP) is used to perform the dynamics evolution.[49, 106-109] The primary idea behind the TDVP is to constrain the time propagation of a quantum state to a specific MPS manifold.[106, 108] It can be understood as the projection of the evolution vector onto the tangent space defined by a given MPS manifold:

$$i\hbar \frac{d}{dt}|\Psi(t)\rangle = \hat{P}_{T,|\Psi(t)\rangle}\hat{H}|\Psi(t)\rangle, \tag{11}$$

where $\Psi(t)$ denotes the total wave function, $\hat{P}_{T,|\Psi(t)\rangle}$ is the projector on the tangent space, which can be decomposed into two terms using the left projector $\hat{P}_i^L$ and right projector $\hat{P}_i^R$:

$$\hat{P}_{T,|\Psi(t)\rangle} = \sum_{i=1}^{m} \hat{P}_{i-1}^L \otimes \hat{I}_i \otimes \hat{P}_{i+1}^R - \sum_{i=1}^{m-1} \hat{P}_i^L \otimes \hat{P}_{i+1}^R, \tag{12}$$

$$\hat{P}_i^L = \sum_{\alpha_i} |\Psi_{L,\alpha_i}^{[1:i]}\rangle \langle \Psi_{L,\alpha_i}^{[1:i]}|, \tag{13}$$

$$\hat{P}_i^R = \sum_{\alpha_i} |\Psi_{R,\alpha_{i-1}}^{[i:m]}\rangle \langle \Psi_{R,\alpha_{i-1}}^{[i:m]}|. \tag{14}$$

In practice, the above equation of motion can be solved approximately by solving $m$ forward-evolving equations and $(m-1)$ backward-evolving equations individually and sequentially:

$$i\hbar \frac{d}{dt}|\Psi(t)\rangle = \sum_{i=1}^{m} \hat{P}_{i-1}^L \otimes \hat{I}_i \otimes \hat{P}_{i+1}^R \hat{H}|\Psi(t)\rangle, \tag{15}$$

$$i\hbar \frac{d}{dt}|\Psi(t)\rangle = -\sum_{i=1}^{m-1} \hat{P}_i^L \otimes \hat{P}_{i+1}^R \hat{H}|\Psi(t)\rangle. \tag{16}$$

The reduced density matrix for the electronic part is given by the following equation

$$\rho_{ij}(t) = Tr\{|\varphi_i\rangle\langle\varphi_j|\Psi(t)\rangle\langle\Psi(t)|\}, \tag{17}$$

where the diagonal ($i = j$) and off-diagonal ($i \neq j$) elements represent the electronic populations and coherences, respectively.

For the initial condition, the lowest vibrational level of the ground state minimum is vertically placed to one LE state. The basis set for the bath mode is determined by the convergence test of the TDVP dynamics propagation. The cutoff value is set as $10^{-13}$ for single value decomposition in the MPS and MPO constructions. The whole dynamics propagation lasts for 1.0 ps with the time step of 0.5 fs.

*2.3. LSTM-RNN.*

As shown in Figure 1(a), the RNN belongs to a class of directional NNs, which receives the outputs of the previous time step and uses them as inputs of the current step. By taking the temporal features into account, the RNN can be used in the prediction of the future evolution. Therefore, the RNNs were widely used in various time-series analysis tasks,[70, 110-113] such as speech recognition and natural language processing.

The essential part of the RNN is the recurrent layer, which corresponds to the function

$$[y_{(t)}, h_{(t)}] = f(h_{(t-1)}, x_{(t)}), \qquad (18)$$

where $x_{(t)}$ is the input at the current time step and $h_{(t-1)}$ is the output given by the previous time step. The recurrent layer gives the output of $y_{(t)}$ and $h_{(t)}$, and they may be the same vector or not.

The LSTM cell is normally used in the recurrent layer, because it largely alleviates gradient vanishing and exploding problems[78] in the RNN training procedure. The topology of LSTM cell is given in Figure 1(b), which includes some control gates.[70, 78]

At time step *t,* the LSTM cell receives three vectors, i.e. $x_{(t)}$ (the input vector), $h_{(t-1)}$ (the short-term state vector) and the $c_{(t-1)}$ (the long-term state vector). All three vectors pass three gates controlled by the *sigmoid* function that is represented as $\sigma$.

(1) Gate 1: the input gate $i_{(t)}$ is defined as

$$i_{(t)} = \sigma(W_{xi}^T x_{(t)} + W_{hi}^T h_{(t-1)} + b_i), \tag{19}$$

which indicates that how much information of the current inputs $x_{(t)}$ and the memory information $h_{(t-1)}$ provided by the previous step should be accepted in the current LSTM cell. Here and the same as below, the tensors $W$ and $b$ are the weight and bias parameters of the LSTM cell. We used different subscripts to label their corresponding gates.

(2) Gate 2: the forget gate $f_{(t)}$ is given as

$$f_{(t)} = \sigma(W_{xf}^T x_{(t)} + W_{hf}^T h_{(t-1)} + b_f), \tag{20}$$

which determines how much input information should be remained.

(3) Gate 3: the output gate $o_{(t)}$ is defined as

$$o_{(t)} = \sigma(W_{xo}^T x_{(t)} + W_{ho}^T h_{(t-1)} + b_o), \tag{21}$$

which decides which information should be saved and passed to the output.

At the same time, an intermediate vector $g_{(t)}$ is generated via a *tanh* activation function as

$$g_{(t)} = tanh(W_{xg}^T x_{(t)} + W_{hg}^T h_{(t-1)} + b_g), \tag{22}$$

which provides the additional controls of the information transfer. Next, the long-term state vector at the current time step $c_{(t)}$ is determined by the output of the forget gate $f_{(t)}$, the long-term state vector from the previous step $c_{(t-1)}$, the input gate $i_{(t)}$ and the intermediate vector $g_{(t)}$, i.e.

$$c_{(t)} = f_{(t)} \cdot c_{(t-1)} + i_{(t)} \cdot g_{(t)}. \tag{23}$$

Finally, the output of the LSTM cell is given as

$$y_{(t)} = h_{(t)} = o_{(t)} \cdot tanh(c_{(t)}), \tag{24}$$

which is the product between the output gate $o_{(t)}$ and the $tanh$ function of the long-term state vector $c_{(t)}$.

The LSTM cell realizes the effective transmission of time series information by above multi-gating mechanisms. It is straightforward to simply stack several LSTM layers to define a multi-layer LSTM-RNN. In the current work, there were several stacked LSTM hidden layers and one dense layer. First, many small time-series fragment sets were passed into the LSTM-RNN model though the input layer. Second, the time-correlated information was analyzed and recorded by the LSTM hidden layers. Finally, the dense layer was employed to extract the information map and to give the result via the output layer.

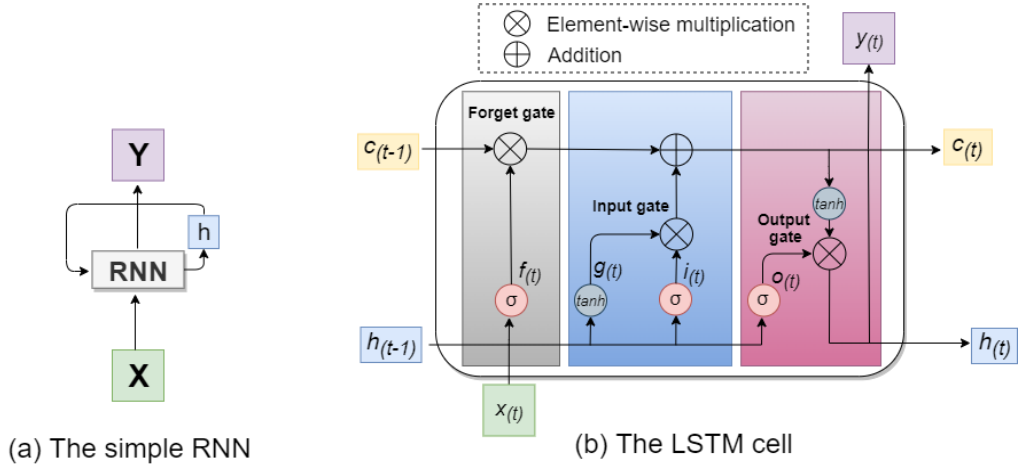

Figure 1. (a) The simple RNN model; (b) The LSTM cell.

*2.4 Hyperparameter Optimization.*

In fact, all parameters in a NN model are divided into two categories. Some parameters that can be obtained directly by the model regression, such as the weight and bias of each neuron, are called regression parameters. The other group of parameters must be defined before the training process, instead of being obtained directly from the regression. In the current work, the number of NN hidden layers, the number of neurons in each layer and the length of the small-piece time-series dataset belong to this category. They are called as hyperparameters[80-81] in the model construction. Therefore, how to select the best hyperparameter sets to achieve the high model accuracy is rather critical in the NN construction.

In the hyperparameter optimization, the most straightforward method is the so-called grid search approach, [114] i.e. taking all possible hyperparameter combinations to define different NN structures, performing the training tasks and finding an optimal NN structure based on the model performance. However, when the network topology becomes complicated, such grid search is not efficient due to high computational cost, as a huge number of hyperparameter combinations must be considered. Thus, the automatic and efficient hyperparameter optimization approaches are highly useful. Here we used three hyperparameter optimization methods given as below.

*2.4.1 Random Search.*

In the RS approach,[87-89, 115] a hyperparameter set is randomly sampled within the user-defined range of the searching space and the chosen one is used in the NN model training. This procedure is repeated until the loss function below certain threshold.

*2.4.2 Bayesian Optimization with Tree-Structured Parzen Estimator.*

The Bayesian optimization method is widely used in the hyperparameter optimization in the NN construction.[85-87, 116-117] As an iterative algorithm, this approach determines the next hyperparameter set according to historical results.

Let us assume that we wish to optimize a scalar objective function $f(x)$, in which $x$ represents the hyperparameter in the current work. Our purpose is to find an optimal solution of $x$, which gives the minimum of the loss function of $f(x)$. With different $x$ values, i.e. $(x_i)$, all possible solutions of $f(x_i)$ should form a distribution. In the Bayesian optimization process, a surrogate model is defined, which represents the probability distribution of the objective function $f(x)$. Several algorithms were proposed to define the surrogate model in the optimization of the objective function $f(x)$, which include Gaussian processes,[86, 118] TPE[86] and so on.

In the current work, the TPE is employed. Since more mathematic information and implementation details were comprehensively discussed in previous works,[86] we only outline the main idea of the BO-TPE approach here. Within the BO-TPE framework, the conditional probability model is built by applying Bayesian rules.

Starting from different hyperparameters $x_i$, the training procedure generates $f(x_i)$. After a set of $\{x_i, f(x_i)\}$ are collected, the BO-TPE optimization divides these observation results into two parts, namely good results and poor results according to a pre-defined percentile threshold $y^*$. For each part, its own individual density function [$l(x)$ or $g(x)$] is determined by the Parzen window as

$$p(x|y) = \begin{cases} l(x), & if\ y < y^* \\ g(x), & if\ y \geq y^* \end{cases}, \qquad (25)$$

where $y$ corresponds to the $f(x)$. It means that the hyperparameter space is divided to two subspaces and their dividing line is determined by the threshold $y^*$. This probability function serves as the surrogate model. Next a new point is sampled by maximizing the so-called acquisition function i.e. expected improvement (*EI*)[119]:

$$EI_{y^*}(x) \propto (\gamma + \frac{g(x)}{l(x)}(1-\gamma))^{-1}, \qquad (26)$$

with

$$\gamma = p(y < y^*) \ \text{and} \ p(x) = \gamma l(x) + (1-\gamma)g(x). \qquad (27)$$

The maximization of the acquisition function gives a new point of $x$ and the inclusion of this point in principle should increase the ratio between the good distribution and the poor distribution. Next, the iterative optimization is performed, in which the algorithm returns the new $x$ ($x_{i+1}$) with the greatest *EI* value. Then, the above TPE model is modified after the calculation of the objective function $f(x_{i+1})$. After repeating the above step many times, the BO-TPE optimization method may find the optimal hyperparameter set.

*2.4.3 Simulated Annealing.*

As a popular hyperparameter optimization algorithm, the SA[82-84] creates a trajectory (*i.e.* $[x_0, x_1, \cdots, x_k]$) by the Metropolis MC strategy to minimize the target loss function $f(x)$. First, the initial state $x_0$ is chosen randomly in the pre-defined searching space. Second, at each step $t$, the random perturbation is applied to the current $x_t$, and this generates a new state $x_{t+1}$. Next whether the $x_{t+1}$ is accepted or not is determined by the probability

$$p = \begin{cases} \exp(-\dfrac{f(x_{t+1}) - f(x_t)}{T}) & \text{if } f(x_{t+1}) - f(x_t) \geq 0 \\ 1 & \text{if } f(x_{t+1}) - f(x_t) < 0 \end{cases}, \quad (28)$$

with the parameter $T > 0$ called "temperature". The mathematic insight of this probability function can be understood as below. When the new state $x_{t+1}$ shows the lower value of $f(x)$, this state is accepted. If the new state shows the higher value of $f(x)$, there is some probability to accept it. By following the above procedure, the SA optimization may give an optimal state $x$.

*2.5 ML Prediction Uncertainty.*

The estimation of the uncertainty is un-avoidable in all ML models. The uncertainty of the ML model is caused by different reasons.[120-121]

- ***Model Misspecification.*** The training and prediction data may follow the different distribution patterns. In this case, the ML model may not describe the feature space spanned by the prediction dataset well.

- ***Model Uncertainty.*** Many NN models may be the available solutions for the same regression dataset. Thus, these models themselves form a distribution, giving the model uncertainty. More precisely, the model uncertainty mainly includes two important key points in the NN models. The first is that different NN models may be given, i.e. the regression may result in several NNs with different structures. The second is that different parameter sets may be obtained for the NN models with the same structure.

All of these uncertainties are normally entangled in realistic situations. Here the origins of these uncertainties themselves give us the key idea on how to estimate them. By invoking the MC treatment, we may generate many different training datasets via resampling from the whole training dataset, train many different NN models and even retrain the NN model several times to get different parameter sets. According to these concepts, two powerful approaches are normally employed to estimate the confidence interval of the ML models, which are bootstrap resampling and MC dropout approaches.

*2.6 Bootstrap Resampling.*

Starting from the training dataset, we may create different data distributions if some data points are removed. The bootstrap algorithm provides a MC description on the fact that different distributions may be formed from the known training dataset. In the bootstrap approach,[73-76] many training datasets are generated and each of them has the same size as the original training dataset. Here, each generated dataset is constructed by random picking an element from the original training set, and thus the same element may appear several times in each new dataset. The idea of the bootstrap resampling procedure is explained in Figure 2. Given an original training dataset shown in Figure 2(a), the bootstrap resampling approach may generate many new sets as shown in Figure 2(b).

In implementation, we first build a primary LSTM-RNN model from the preliminary training step. Then the topology and connectivity of the LSTM-RNN model is fixed. Starting from it, several independent LSTM-RNN models with the fixed topology is re-trained based on each resampled dataset and all parameters are re-fitted. Clearly, the bootstrap approach naturally includes different distribution patterns spanned by the training data, giving the MC description on the model misspecification. Since all LSTM-RNN models display different parameters, the bootstrap approach also takes the parameter uncertainty into account. When several LSTM-RNN structures are considered in the bootstrap, we can also access the ML model distribution partially.

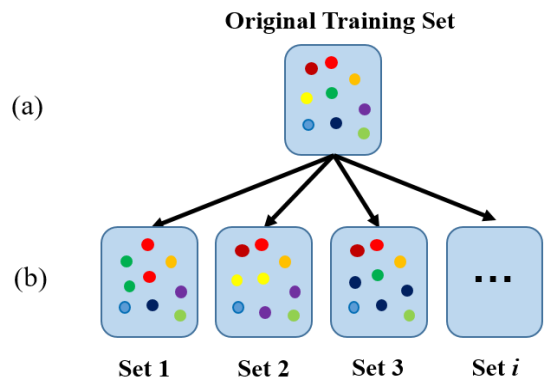

Figure 2. The bootstrap resampling process: performing the random data resampling from the original training dataset (a) to generate many new training datasets (b). Here each new training dataset has the same data size and contains duplicated elements. Each colour represents an individual data point in the original dataset.

*2.9 Monte-Carlo Dropout.*

The MC dropout suggested by Gal et. al[101-102] is another widely-used approach to estimate the uncertainty of the NN model prediction. In fact, originally the standard dropout trick was proposed to avoid over-fitting problems,[78, 122] in which some neurons in a NN are randomly switches off. This can be viewed as a practical way to add the regulation in the NN regression.[101-102]

In the MC dropout approach, the dropout operation is conducted many times and this generates many NN configurations. As shown in Figure 3, each dropout configuration is generated by randomly switching some neurons off (grey circles with cross) or not (blue circles). Finally, an ensemble of all dropout-configuration networks can be used to estimate both the NN model prediction and the confidence interval.

Gal et. al[101-102] once pointed out that the employment of the MC dropout in NNs can be viewed as the stochastic realization of the Bayesian estimation of the NN model uncertainty. More discussions on the rigorous mathematic view can be found in the references.[101-102] Here, a set of NN models with different NN structures and different parameter sets can be built by the MC dropout. In this sense, both model distribution and parameter distribution can be well captured by the MC dropout.

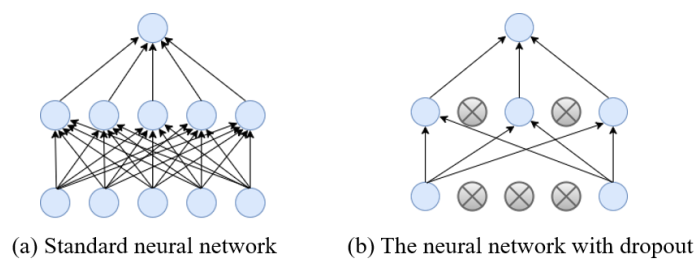

(a) Standard neural network    (b) The neural network with dropout

Figure 3. The structure diagram of standard neural network (a) and its modification with dropout (b).

## III. Computational details

(a) The input of the ML model.

As the reduced density matrix includes the complex number, we represented their elements by a three-dimensional vector, i.e. $X = [\Delta = \rho_{11} - \rho_{12}, \text{Re}\{\rho_{12}\}, \text{Im}\{\rho_{12}\}]$. The short-time evolution of this vector $X$ was interpreted by the LSTM-RNN model construction and the prediction of the ML model gives the evolution of this vector. As a complimentary study, we also try to build the ML model by using only the population difference term $\Delta$. In this way, we may examine the role of the off-diagonal elements in the ML model construction and prediction.

(b) Dataset Preparation.

Let us consider the short-time propagation of the open quantum dynamics up to a time duration. The evolution of the density matrix is represented by a series of the time-dependent vector $[X(t_1), \cdots, X(t_n)]$ with the discretized time step of 0.5 fs. Therefore, we have a time series and the total number of time step is $n$.

Starting from $[X(t_1), \cdots, X(t_n)]$, we may build many time-series data subsets $\{S_i\}$ with the length $L$, as shown in Figure 4. Each subset includes $L$ successive time steps and the time duration of this small time-series set ($0.5*L$ fs) is relevant to the so-called memory time used in some references.[69]

All sequences $\{S_i\}$ were divided into two groups with the ratio of 3:1 according to the chronological order. The early-time sequences ($\{S_i^{(A)}\}$: Group A in Figure 4) was taken to construct the LSTM-RNN model, and the later-time sequences ($\{S_i^{(B)}\}$: Group B in Figure 4) was employed to examine the ML model prediction accuracy.

After all $\{S_i^{(A)}\}$ were randomly shuffled with their individual chronological order unchanged, they were randomly divided into two sub-datasets $[\{S_i^{(A1)}\}, \{S_i^{(A2)}\}]$ with a ratio of 7:3. The first sub-dataset $\{S_i^{(A1)}\}$ defines the training set in the regression of the LSTM-RNN model. The second sub-dataset $\{S_i^{(A2)}\}$ defines the internal validation set that is used for the early-stop training step. The later-time sequences $\{S_i^{(B)}\}$ defines the external validation set used for the selection of the suitable LSTM-RNN model.

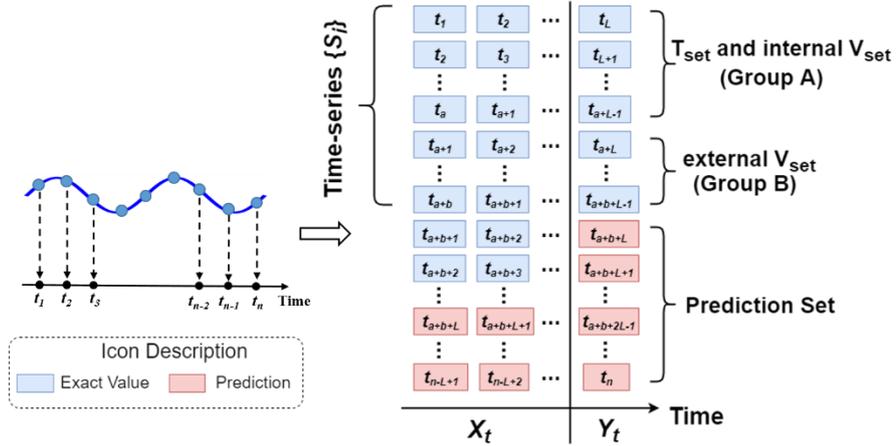

Figure 4. A brief scheme of the time-series dataset slice.

(c) ML model training and validation.

The LSTM-RNN model was trained based on $\{S_i^{(A1)}\}$, in which the mean square error (MSE) of the loss function is minimized by using the adaptive moment estimation (Adam) method. The early stop approach was taken to avoid the overfitting in the basis of $\{S_i^{(A2)}\}$.

In the training step, we need to choose the optimal hyperparameters, including the NN layer number, the neuron number in each layer, the length $L$ of the datasets $\{S_i\}$. Three hyperparameter optimization approaches were taken, which include RS, BO-TPE and SA. The search space spanned by all hyperparameters is defined as follows. The possible number of NN layer is 2, 3 or 4. For the neuron number in each layer, the search is performed within the range of [10, 500] with the interval as 20. Since it is not clear on how to set the proper length of the memory time, we also treated it as the hyperparameter and the optimization search range is from 5 fs to $(L/4)$ fs. In this way, the hyperparameter searching space is defined.

Three hyperparameters optimization methods were employed to conduct the hyperparameter optimization. In this process, the loss function of the NN model in the basis of the external validation set $\{S_i^{(B)}\}$ was considered. In addition to these, the batch size and the epoch number in all LSTM-RNN training tasks are 50 and 300, respectively. In each hyperparameter optimization approach, we totally ran 20 independent optimization tasks starting from different initial conditions. In each optimization task, 100 iterations were taken and the best hyperparameter set is chosen. At the end, we either selected a few of hyperparameter sets or the best one from all 20 jobs to conduct the further prediction. All the hyperparameter optimizations were conducted by using the Hyperopt[123-124] program library and the optimization setups (such as the temperature in SA) simply follows default values in the Hyperopt program.

(d) Prediction Uncertainty.

After the hyperparameter optimization, we obtained several optimal LSTM-RNN models with different NN structures/parameters and their corresponding memory time (length $L$). The next task is to predict the model uncertainty by using two approaches, bootstrap resampling and the MC dropout methods.

We first considered the bootstrap resampling approach, in which the NN structure obtained at the previous step is fixed and only the parameters are adjusted by the regression based on each resampled dataset. Here, the training set and the internal validation set $[\{S_i^{(A1)}\}, \{S_i^{(A2)}\}]$ were mixed together and resampled to generate the new datasets to perform the bootstrap resampling analysis. After this re-construction, a new LSTM-RNN network with the same network structure was re-trained based on a resampled dataset. And the additional parameters including the epoch number and the batch size, are same as the those in the hyperparameter optimizations. We repeated this resampling-regression process many times. For a given LSTM-RNN structure, a group of 100 LSTM-RNNs were obtained in the bootstrap step at the end. To address the model uncertainty caused by different network structures, we selected 10 proper LSTM-RNN structures from the 20 hyperparameter optimization tasks. This gives us totally 1000 (100×10) LSTM-RNN models and all of them were used in the prediction.

To clarify which hyperparameter optimization and uncertainty estimation approaches are selected, we try to use the below labels, such as (SA/BO/RS-H10)× BT100, for illustration. In this given example, the first part (SA/BO/RS-H10) defines the hyperparameter optimization approach. Here, the SA, BO and RS represent the simulated annealing, the BO-TPE optimization and the random search method respectively. H10 means that 10 NN structures are chosen from many independent hyperparameter optimization tasks. And then, the term of BT100 means that each NN structure is re-trained based on 100 times of bootstrap resampling. In the given case, we totally used 1000 (10×100) LSTM-RNN models. The average of the prediction values computed by these LSTM-RNN models gives the future time-series data to characterize the long-time quantum evolution, while the standard deviation of all foresting values defines the model confidence interval.

In the MC dropout approach, we first chose the best LSTM-RNN model (only one) after the hyperparameter optimization. Next, we enlarged the last LSTM layer of the network by doubling the number of the neurons, while all other layers remain unchanged. Following the idea of MC dropout, we randomly switched off 50% neurons in the last LSTM layer to obtain many new LSTM-RNNs. Starting from each randomly-generated NN, the training task was performed, and finally many NN models were obtained to give the further dynamics evolution and to estimate the model uncertainty. The way to perform the dropout task only on the last layer was recommended in previous work[125] due to its good performance.

Similar to the bootstrap case, the labels, such as (SA/BO/RS-H1)×MC100, are employed in the below discussions. Here the first part (SA/BO/RS-H1) defines which hyperparameter optimization approach was taken and how many basic NN models were chosen here. Since the MC dropout step generates many NN models with different structures, the model distribution itself is naturally considered. Thus, we only chose the best NN model (labelled as H1) in the current example to conduct the MC dropout. At the end, we totally obtained an ensemble of the LSTM-RNN models with 100 different network structures.

In principle, the MC dropout approach gives a more completed description on the model uncertainty than the bootstrap resampling approach, since more NN models with different topologies are generated in the former one. However, the bootstrap resampling approach gives a more suitable description on the model misspecification by resampling the training data. Thus, it is also recommended to combine both for the more comprehensive estimation of the LSTM-RNN prediction uncertainty. For illustration, we used the labels (SA/BO/RS-H1)×BT50×MC50 to refer such models. In the given example, the best LSTM-RNN model was chosen in the hyperparameter optimization tasks. Next the new 50 training datasets were built by the bootstrap resampling. For each resampled training dataset, we conducted the MC dropout and obtained 50 different MC dropout NN network structures. Finally, for this given example, we obtained an ensemble of the LSTM-RNN models with 2500 (50×50) NNs. In this way, we combined both bootstrap and MC dropout approaches to give a comprehensive estimation of both model uncertainty and model misspecification.

## IV. Results

In the current work, the preliminary LSTM-RNN models were constructed by using three hyperparameters optimization approaches including SA, BO-TPE and RS. After the selection of a few optimized LSTM-RNN structures, we built several LSTM-RNNs by using the bootstrap resampling approach, the MC dropout method and their combination. Finally, all NNs were taken to perform the prediction and estimate the confidence interval of the prediction.

The first task of this work is to examine the performance of hyperparameter optimization methods. In this step, we chose the symmetric site-exciton model (labelled as Model I) and built the appropriate LSTM-RNN models based on the quantum dynamics up to 350 fs.

*4.1 Hyperparameter Optimization.*

*4.1.1 Error Distribution in Hyperparameter Optimization.*

It is necessary to know which part of the hyperparameter space is travelled in the hyperparameter optimization processes to get the initial idea in the analysis of the performance of three optimization methods. For each of them, we considered 20 independent optimization tasks with 100 steps and this covers 2000 points in the hyperparameter spaces. Next, we calculated the prediction errors of these models based on $\{S_i^{(B)}\}$ and Figure 5(a) shows the error distribution. The SA method tends to give more snapshots with the rather small error, while the BO-TPE and RS approaches experience more large-error points in the hyperparameter space. In fact, the different performances among three methods is larger than their first view impression given in Figure 5(a) due to the employment of the exponential scaling of the x-axis. Such comparison implies that the SA method may find the optimal NN models more easily than other two approaches. This idea is confirmed by examining the performance of the final optimal LSTM-RNN models. Since each optimization task gives a single LSTM-RNN model at the end, we totally obtained 20 models whose topologies are given in Table S2 in SI. Here we sorted all models from 1 to 20 according to their prediction errors in the basis of the external validation dataset $\{S_i^{(B)}\}$, as shown in Figure 5(b). It is clearly that the SA method still gives best results. The performances of other two methods are not far from each other, while the RS approach gives the slightly better LSTM-RNN models than the BO-TPE method.

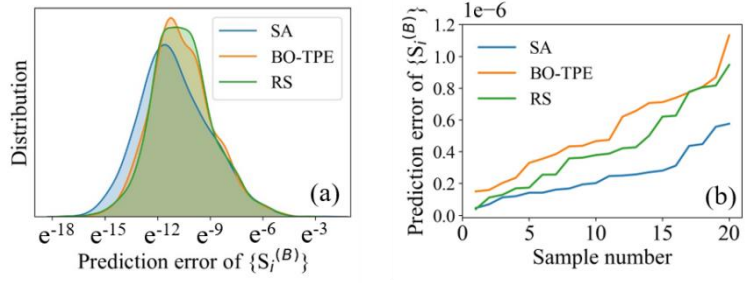

Figure 5. (a) The distribution of prediction errors of all involved LSTM-RNN models in the hyperparameter optimization steps with three methods (each method covers 2000 snapshots); (b) The validation errors of the optimal LSTM-RNN models obtained from 20 independent hyperparameters optimization tasks. The validation error is defined in the basis of the external validation set $\{S_i^{(B)}\}$. And the SA, BO, RS represent the simulated annealing method, Bayesian optimization method with tree-structured parzen estimator and the random search approach, respectively.

*4.1.2 The Distribution of Memory Time.*

The reasonable prediction of the dynamical evolution of the open quantum systems at the long-time scale is based on whether the dynamical map built from the historical short-time dynamics can capture the dynamical correlation or not. When the historical dynamics length is given, we need to prepare many small time-series data (Figure 4) to define the input for the LSTM-RNN network. This short-time duration is called "memory time" in the previous works.[69] Since we do not know how to choose its proper value, it is a hyperparameter. In this work, its optimal value is also automatically determined by the hypeparameter optimization approaches. As each optimization method gives 20 values, we simply collected all and showed their distribution in Figure 6. It is clearly that the highest peak in the current distribution is located at ~ 53 fs and the whole peak range is ~ 35-71 fs. We wish to emphasize that the current hyperparameter optimization approaches not only give us the optimal LSTM-RNN structures, but also provide the proper memory time automatically. This largely improves the applicability of the LSTM-RNN models in the simulation of the quantum dissipative dynamics.

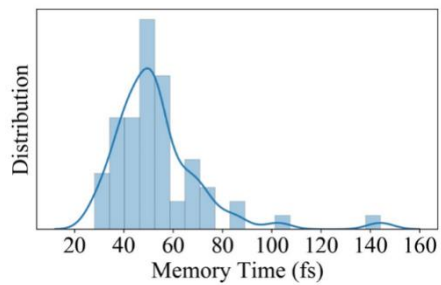

Figure 6. The distribution of memory time obtained by all three hyperparameter optimization methods.

*4.2 Combinations of Hyperparameter Optimization and Uncertainty Estimation.*

*4.2.1 The LSTM-RNN Prediction with Bootstrap.*

We examined the LSTM-RNN prediction of the quantum dynamics and analyzed the model confidence interval with the bootstrap approach. The (SA/BO/RS-H10)×BT100 LSTM-RNN models were employed to perform the dynamics simulation and the results are given in Figure 7. At the first glance, the long-time quantum dynamics is well predicted no matter which hyperparameter optimization approach is taken. For better view, the prediction deviation and the confidence interval are displayed as the functions of time being in Figure 8. The prediction error is defined by the difference between the LSTM-RNN model prediction values and the exact values obtained by the tensor-train dynamics simulation, while the confidence interval given in the bootstrap resampling procedure characterizes the prediction uncertainty. All hyperparameter optimization approaches give the similar prediction errors for the dynamics data and all deviations are very small. As expected, the deviations become larger in the later-stage dynamics, while they are still acceptable. Overall, the bootstrap-based LSTM-RNN models give the reasonable prediction of the quantum evolution. In convergence tests, we also considered (SA/BO/RS-H5)×BT100 and (SA/BO/RS-H1) × BT100 LSTM-RNN models and the results are given in Figure S1-S4 in SI, respectively. It is indicated that the (SA/BO/RS-H10)×BT100 prediction results is convergent.

We notice that the SA and BO-TPE approaches seem to show the better performance than the RS method after the inclusion of the bootstrap. For instance, the RS optimization method seems to give the larger uncertainty in the prediction of the long-time dynamics (Figure 8). The different performance of the SA and RS approaches may be attributed to the fact that the former one generally results in the better LSTM-RNN models as shown in Figure 5(b). Here the BO-TPE method seems to give the better performance than the RS method (Figure 8), although the optimal LSTM-RNN models given by BO-TPE method themselves do not display the smaller prediction errors (Figure 5(b)). The analysis of the underline reasons may require the additional works in the future. In the current work, it is enough to access their performance only from numerical results.

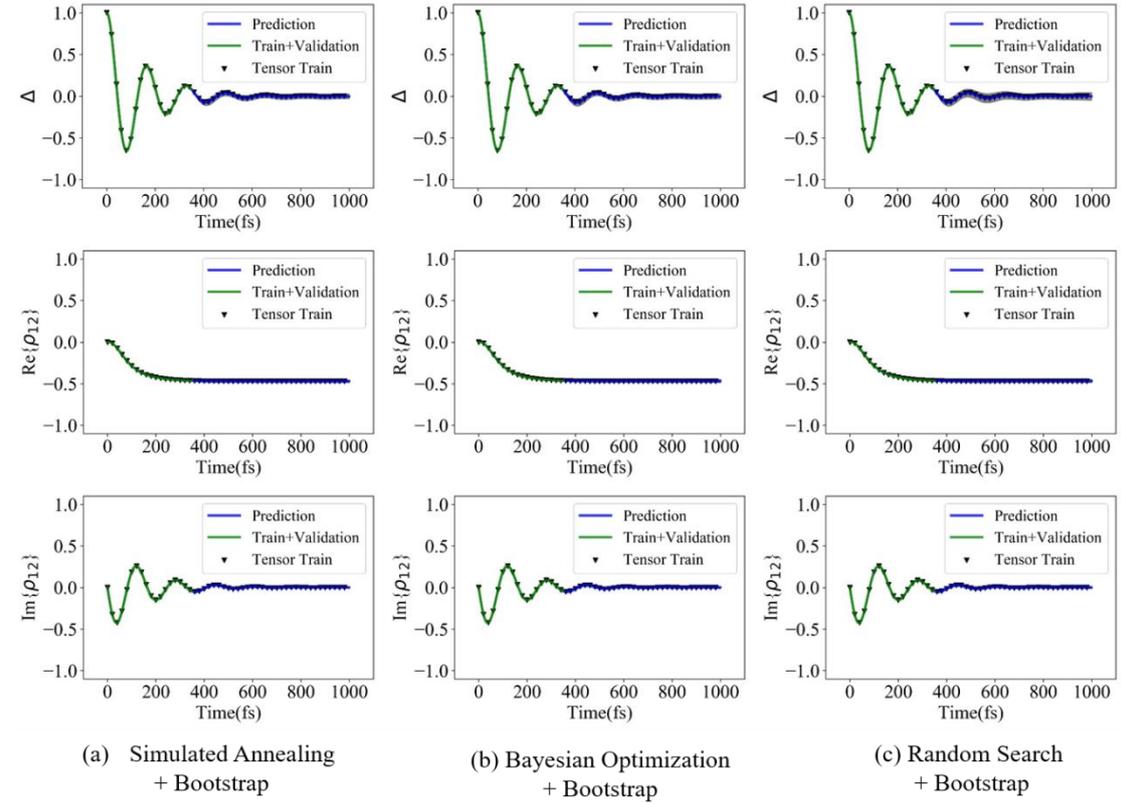

Figure 7. The quantum dynamics simulated by the (SA-H10)×BT100 (a), (BO-H10)×BT100 (b), and (RS-H10)×BT100 (c) LSTM-RNN models *vs.* the tensor-train quantum propagation in Model I. The green lines denote the training and validation samples (<350 fs) used in the LSTM-RNN model construction. The black triangles display the tensor-train simulation results. The blue lines correspond to the LSTM-RNN prediction of the future dynamics and the grey region shows the prediction uncertainty.

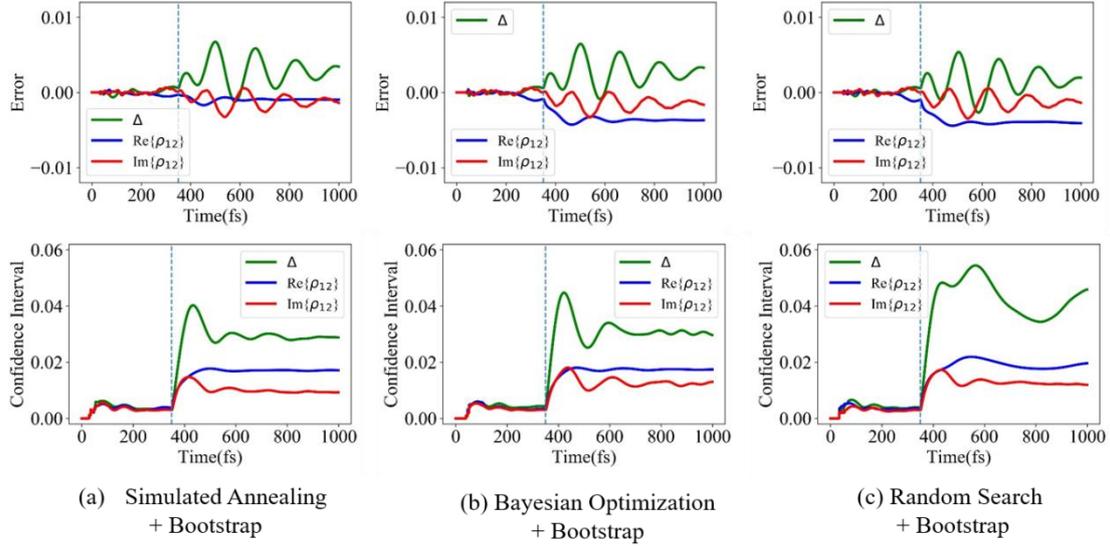

Figure 8. The prediction error and confidence interval of the (SA-H10)×BT100 (a), (BO-H10) ×BT100 (b), and (RS-H10)×BT100 (c) LSTM-RNN simulation of the quantum dynamics of Model I. The blue dotted lines denote the time duration (< 350 fs), in which the training and validation samples were employed in the LSTM-RNN model construction.

*4.2.2 The LSTM-RNN Prediction with MC Dropout.*

When the MC dropout method was taken to estimate the model uncertainty, all LSTM-RNN fitting and prediction results, along with the quantum dynamics data, are displayed in Figure 9. Three hyperparameter optimization methods show rather different performances. The SA approach behaves quite well, while the large confidence interval exists in the RS approach. All details of the prediction error and uncertainty are given in Figure 10, in which the RS results are not included due to its poor performance (see Figure S5 in SI). According to Figure 9 and 10, the selection of three hyperparameter optimization methods follows the order as: SA > BO-TPE > RS. The underline reason of the excellent performance of the SA approach was discussed in the previous section, *i.e.* its optimization pathway always tends to give new points with lower prediction error.

In the MC dropout approach, a large number of the NN models with different structures were generated. This may explain that the prediction uncertainty becomes larger in the MC dropout approach.

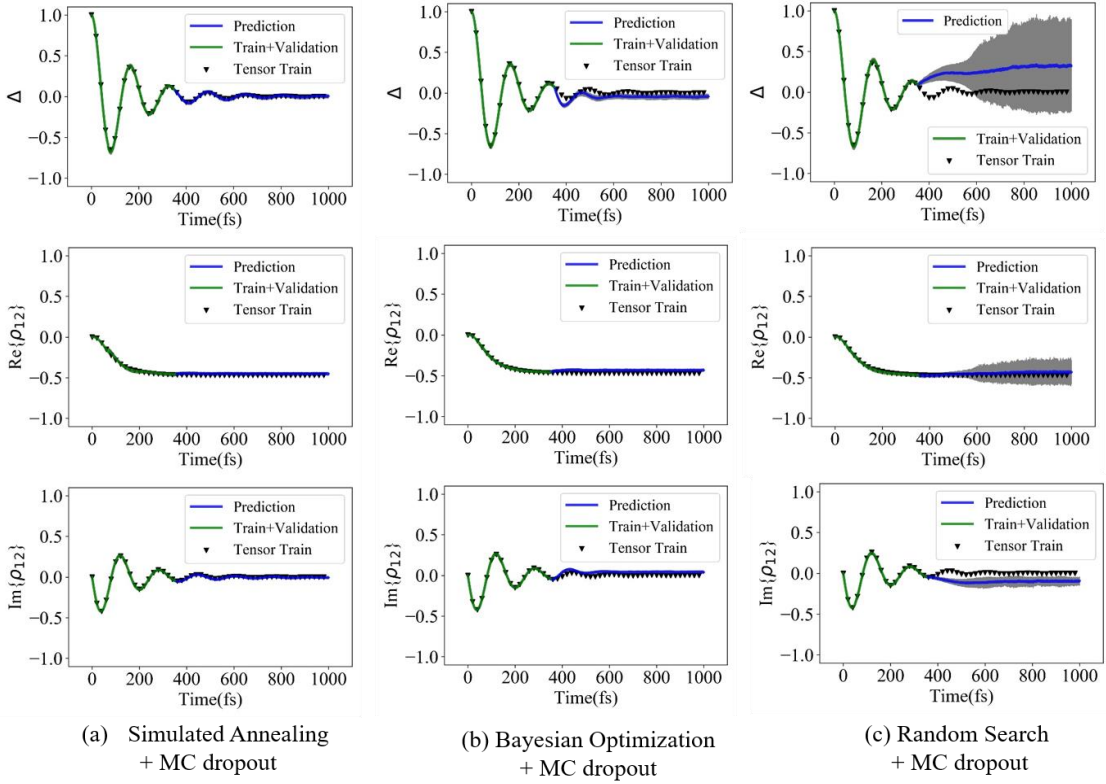

(a) Simulated Annealing + MC dropout
(b) Bayesian Optimization + MC dropout
(c) Random Search + MC dropout

Figure 9. The quantum dynamics simulated by the (SA-H1)×MC100 (a), (BO-H1)×MC100 (b), and (RS-H1)×MC100 (c) LSTM-RNN models *vs.* the tensor-train quantum propagation in Model I. The green lines denote the training and validation samples (<350 fs) used in the LSTM-RNN model construction. The black triangles display the tensor-train simulation results. The blue lines correspond to the LSTM-RNN prediction of the future dynamics and the grey region shows the prediction uncertainty.

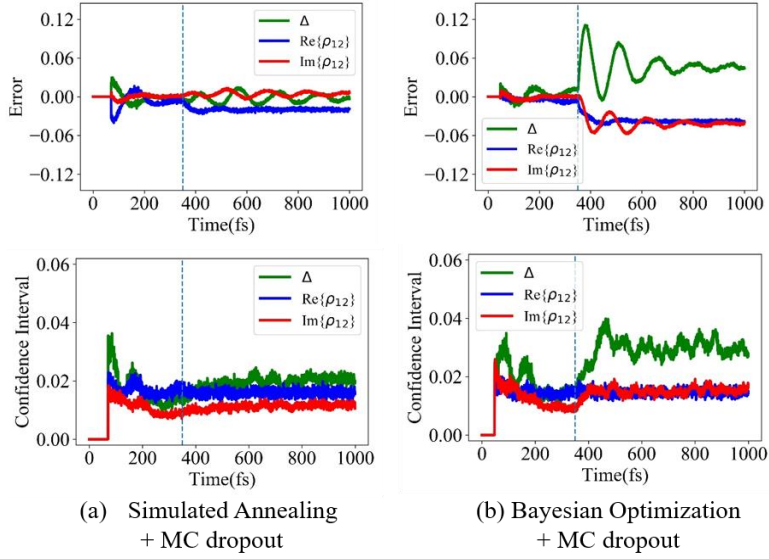

Figure 10. The prediction error and confidence interval of the (SA-H1)×MC100 (a) and (BO-H1)×MC100 (b) LSTM-RNN simulation of the quantum dynamics of Model I. The blue dotted lines denote the time duration (< 350 fs), in which the training and validation samples were employed in the LSTM-RNN model construction.

*4.2.3 The LSTM-RNN Prediction by Combining Bootstrap and MC Dropout.*

Clearly, the bootstrap and MC dropout methods try to address different prediction uncertainties. In order to get the full estimation of the whole uncertainty, we tried to apply both methods simultaneously. Similar approach was employed in previous work.[121] In the current implementation (SA/BO/RS-H1)×BT50×MC50, the combination of the bootstrap and MC dropout methods totally gives 2500 LSTM-RNNs (1 LSTM-NN model from the hyperparameter optimization, 50 bootstrap resampling operations and 50 MC dropout operations). All results are summarized in Figure 11 and 12. To make sure that the convergence is reached, we randomly selected 1000 LSTM-RNN models from them and ran the prediction, see Figure S6 and Figure S7 in SI.

Here, the SA method is still the best choice, in which the reliable and stable prediction of the quantum evolution is achieved. For the BO-TPE method, the sudden increasing of the prediction uncertainty is observed in the later-stage of the quantum propagation, implying the existence of the instable prediction for the long-time dynamics. For the RS + bootstrap + MC dropout approach, both the prediction error and uncertainty become smaller with respect to the RS + MC dropout approach, possibly due to the error cancelation. Overall, the SA + bootstrap + MC dropout combination seems to be a recommended choice in practices. In this way, the SA approach builds the preliminary proper LSTM-RNN model, and then both bootstrap and MC dropout methods are employed to estimate the model uncertainty and misspecification at the same time.

However, the combination of the bootstrap and MC dropout methods needs very large computational cost, as many NN models are built. Therefore, to reduce the computational efforts in implementation, we recommend to use the bootstrap approach solely for a rough estimation of the prediction uncertainty. Although the bootstrap cannot fully address the distribution formed by many NN models with different connectivity, it can be realized quickly. At the same time, we may use several NN models with different structures in the bootstrap step to partially consider the model uncertainty. For this purpose, we may follow the below procedure. First let us use the bootstrap to get a preliminary and rapid understanding of the model stability. When such uncertainty is small, it indicates that the optimal NN model is obtained. It is worthwhile to combine both bootstrap resampling and MC dropout approaches to give a full estimation of the model uncertainty. Otherwise, the NN model prediction becomes not stable and we should not trust its modelling results any more. This protocol may save considerable computational time in the estimation of the NN model uncertainty.

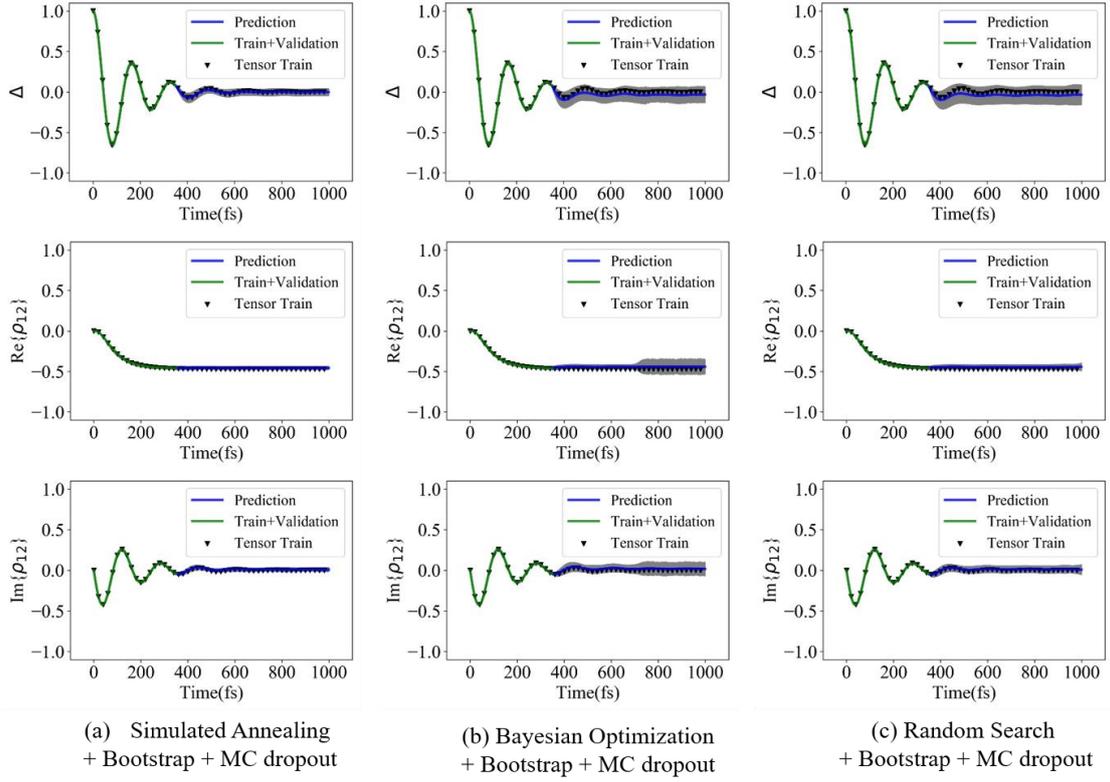

(a) Simulated Annealing
+ Bootstrap + MC dropout

(b) Bayesian Optimization
+ Bootstrap + MC dropout

(c) Random Search
+ Bootstrap + MC dropout

Figure 11. The quantum dynamics simulated by the (SA-H1)×BT50×MC50 (a), (BO-H1)× BT150×MC50 (b), and (RS-H1)×BT50×MC50 (c) LSTM-RNN models *vs.* the tensor-train quantum propagation in Model I. The green lines denote the training and validation samples (<350 fs) used in the LSTM-RNN model construction. The black triangles display the tensor-train simulation results. The blue lines correspond to the LSTM-RNN prediction of the future dynamics and the grey region shows the prediction uncertainty.

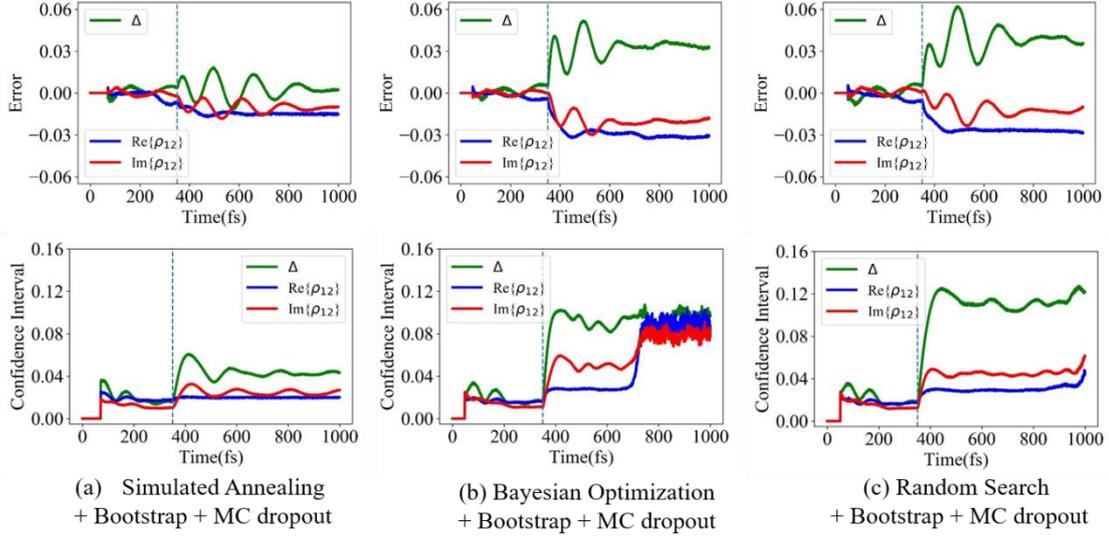

Figure 12. The prediction error and confidence interval of the (SA-H1)×BT50×MC50 (a), (BO-H1)×BT50×MC50 (b) and (RS-H1)×BT50×MC50 LSTM-RNN simulation of the quantum dynamics of Model I. The blue dotted lines denote the time duration (< 350 fs), in which the training and validation samples were employed in the LSTM-RNN model construction.

## 4.3 The Prediction based on Different Time-Series Training Data.

### 4.3.1 The Prediction based on Shorter Time-Series Training Data.

In this part, we examined the performance of the LSTM-RNN models when the shorter dynamics evolution data (~200 fs) was used in the model construction. We only applied the SA + bootstrap approach to obtain the quick view of the prediction quality.

Compared with the results obtained by the LSTM-RNN models constructed based on the 350 fs time-length quantum propagation data, the current prediction accuracy and reliability become lower (Figure 13). This indicates that the performance of the LSTM-RNN models is highly dependent on the length of the time-series data in the training dataset. In principle, this time duration must be long enough to capture the dynamical features in the quantum evolution. However, this brings additional difficulties in the realization of the current theoretical approaches, since we do not have the clear idea on how long the short-time dynamics data should be taken in the model construction. Here we tried to propose two practical approaches to solve this problem.

The first approach is suitable for the situation when our purpose is to simulate the quantum dynamics up to a pre-defined time duration, for instance 1 ps. In realization, we can simply build the NN models based on several sets of time-series training data with different historical lengths, such as 100 fs, 200 fs, 300 fs and so on, and then use them to simulate the long-time dynamics to 1 ps. When the time length is larger than a threshold, the prediction results remain unchanged and the convergence is achieved. This provides a practical way to determine the suitable time duration for the whole training data length.

The second approach works, if only a short-time dynamics data is given. In this case, the prediction uncertainty gives a preliminary judgement on the model reliability, although they are not fully equivalent. In practices, we build an ensemble of NN models to estimate the prediction uncertainty. When the confidence interval becomes larger than the pre-defined threshold with time being, the NN models become not trustable anymore and we should stop to use it for the quantum propagation. This provides an alternative way to know the reliability of the long-time propagation with the built LSTM-RNN models.

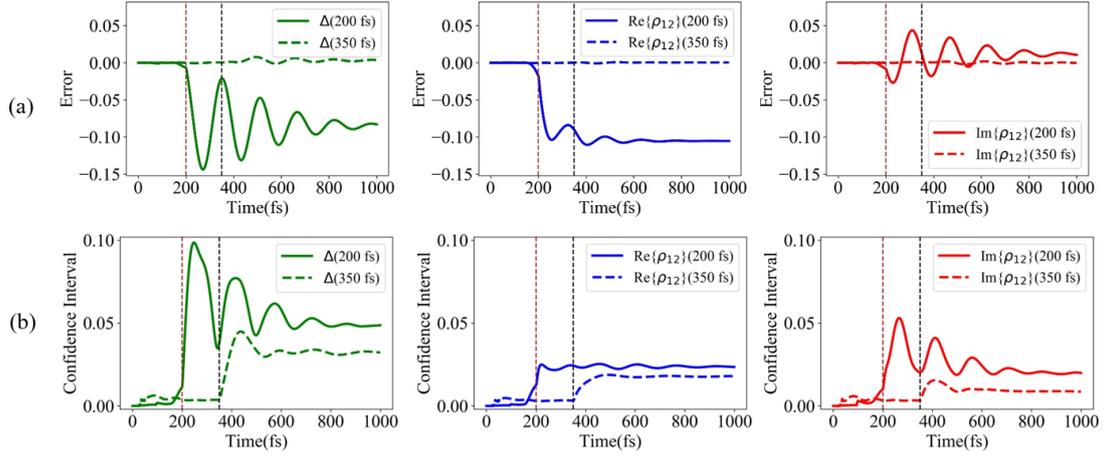

Figure 13. Prediction error (a) and confidence interval (b) with different short-time dynamics data. Here the (SA-H10)×BT100 bootstrap-based LSTM-RNN model is used in the simulation of the quantum dynamics of Model I. The brown and black vertical dotted lines indicate the time durations of 200 fs and 350 fs, respectively.

*4.3.2 The Prediction only based on the Electronic Populations.*

The time-dependent electronic populations themselves provide the very primary understanding of the quantum dynamics of reduced systems. We wish to check the performance of the LSTM-RNN models, when only the population difference term was taken in the model construction. To obtain the rough idea, the SA + bootstrap approach was used and the results are given in Figure 14. Although the overall prediction looks fine, the prediction uncertainty becomes extremely high and prediction is not stable. Therefore, it is not recommended to only include the diagonal elements of the reduced density matrix in the LSTM-RNN model construction. This conclusion is highly consistent with the physical insight: the off-diagonal elements of the density matrix, which characterize the quantum coherence, should play a very important role in the quantum dynamics evolution of reduced systems.

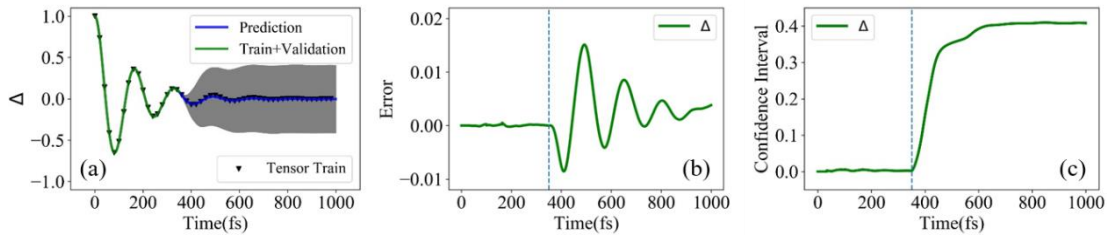

Figure 14. (a) The quantum dynamics modelled by the (SA-H10)×BT100 bootstrap-based LSTM-RNNs *vs.* the tensor-train quantum dynamics of Model I. The green lines denote the training and validation samples used in the LSTM-RNN construction as shown as 350 fs. The blue lines correspond to the LSTM-RNN prediction of the future dynamics. The black triangles display the tensor-train simulation results. And the grey region represents the confidence interval of the forecasting results. The prediction error and uncertainty are shown as (b) and (c) respectively.

*4.4 The Prediction the Quantum Evolution of Other Site-Exciton Models.*

Following the (SA-H1)×BT50×MC50 procedure, we simulated the quantum evolution of other site-exciton models, which includes both symmetric and asymmetric models with different system-bath coupling strengths. They were labelled as Model II, Model III and Model IV respectively with the different system-bath coupling parameters given in Table S1 in SI. All dynamics results are shown in Figure 15 and more details are given in Figure S8 in SI. Overall, the LSTM-RNN models can give the reasonable description of the long-term quantum evolution for these models, while the simulation accuracy is still system dependent. For the symmetric site-exciton models, we always obtained the very reliable results: the prediction result is very accurate and the NN model uncertainty is very low. For the asymmetric site-exciton models, we still can obtain the correct evolution dynamics while the confidence interval becomes visible. The prediction results by the LSTM-RNN models are not very stable now. This indicates that the current LSTM-RNN approach may work better in the symmetric site-exciton models rather than in the asymmetric site-exciton models. In fact, the previous work by Wu et. al. [28] also found the similar features. They argued that the prediction accuracy in the dynamics evolution of the asymmetric system-plus-bath model may be significantly improved by using the transfer learning approaches. This gives us a good inspiration to improve our method in the future.

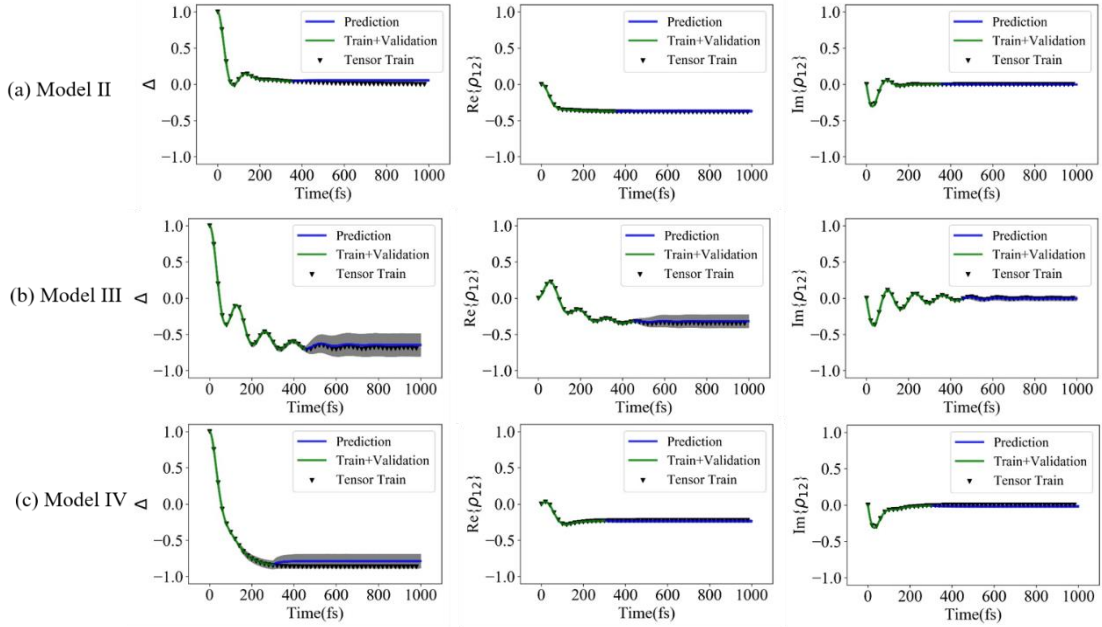

Figure 15. The quantum dynamics modelled by the (SA-H1)×BT50×MC50 LSTM-NN models *vs.* the tensor-train quantum propagations in Model II (a); Model III (b) and Model IV (c). The green lines denote the training and validation samples used in the LSTM-RNN construction. The black triangles display the tensor-train simulation results. The blue lines correspond to the LSTM-RNN prediction of the future dynamics and the grey region shows the prediction uncertainty.

## V. Conclusion

In this work, we tried to employ the LSTM-RNN models to simulate the dynamics evolution of the open quantum systems. The LSTM-RNN models were built based on the short-time historical dynamics evolution data. If such models can well capture the dynamical correlation, we may use the obtained LSTM-RNN model to simulate the long-term quantum dynamics effectively.

We tried to employ three automatic approaches to perform the hyperparameter optimization in the model construction, which are SA, BO-TPE and RS. After the hyperparameter optimization, the optimal NN structures and other corresponding hyperparameters are given. Among these approaches, the behaviour of the SA method is always excellent and thus this is recommended in the future works.

Two approaches were considered to estimate the model prediction uncertainty, which are the bootstrap resampling and MC dropout approaches. In this sense, we provided not only the simulation results of the quantum evolution itself but also the prediction uncertainty. The bootstrap can well describe the model misspecification, *i.e.* the training and prediction data do not follow the same distribution. It can also partially provide the model uncertainty caused by the distribution of the fitting parameters. By using several NN structures, the bootstrap method can also take the NN model distribution into account. The MC dropout method can address two types of model uncertainties very well, what are relevant to the NN model distribution and the fitting parameter distribution. The combination of both two approaches can provide the full assessment of the confidence interval of the prediction. However, such combination may require a rather large computational cost. Thus, it is also recommended to use only the bootstrap resampling approach due to its simplicity and fastness, if only the preliminary estimation on the reliability of the model prediction is needed.

The dynamics simulation by the LSTM-RNN model is also highly dependent on how long the historical data is given in the model construction. Only when this early-time duration is long enough to capture the essential dynamical features, the reliable LSTM-RNN models may be constructed to predict the further long-time dynamics correctly. In the LSTM-RNN model construction, we also realized that it is important to include the off-diagonal elements of the reduced density matrix as well, possibly due to their important roles in the description of the quantum coherence critical to the dynamics evolution of open quantum systems.

Overall, we recommended to combine the SA hyperparameter optimization to build the optimal LSTM-RNN models and the bootstrap + MC dropout approach to perform the prediction uncertainty analysis. This combined approach allows us to build the proper LSTM-RNN models efficiently from the early-stage short-time dynamics and to use them in the simulation of the late-stage long-time quantum evolution effectively. Beside it, we also get the primary idea on the reliability of the quantum dynamics given by the LSTM-RNN models based on the prediction uncertainty. This indicates that the current approach is a feasible approach in the future studies of the quantum evolution of complex systems. In addition, this work discusses several useful ML technical tricks, including hyperparameter optimization methods and uncertainty estimation approaches, as well as their performances. Therefore, we expect that the current work should be very helpful to the future researches that wish to apply the ML models to study other physical and chemical problems.

## VI. Supporting Information Available

Several relevant information: more details of the site-exciton model Hamiltonian; the optimized LSTM-NN model structures in the hyperparameter optimization; the convergence tests of the bootstrap-based LSTM-RNN models; the error and uncertainty analyses of the RS + MC dropout LSTM-RNN models; the convergence tests of the bootstrap + MC dropout LSTM-RNN models; the additional results of other site-exciton models.

## VII. Author Information

**Corresponding Author**

E-mail: gu@scnu.edu.cn; zhenggang.lan@m.scnu.edu.cn.

**Notes**

The authors declare no competing financial interest.

**Acknowledgments**

The authors express sincerely thank to the National Natural Science Foundation of China (No. 21873112, 21933011 and 21903030) for financial support. Some calculations in this paper were done on SunRising-1 computing environment in Supercomputing Center, Computer Network Information Center, CAS.

**Supporting Information for**

**Automatic Evolution of Machine-Learning based Quantum Dynamics with Uncertainty Analysis**

Kunni Lin[1,2], Jiawei Peng[1,2], Chao Xu[2,3], Feng Long Gu[2,3,*] and Zhenggang Lan[2,3,*]

*1 School of Chemistry, South China Normal University, Guangzhou 510006, P. R. China.*

*2 MOE Key Laboratory of Environmental Theoretical Chemistry, South China Normal University, Guangzhou 510006, P. R. China.*

*3 SCNU Environmental Research Institute, Guangdong Provincial Key Laboratory of Chemical Pollution and Environmental Safety, School of Environment, South China Normal University, Guangzhou 510006, P. R. China.*

*\* Corresponding Author. E-mail: gu@scnu.edu.cn; zhenggang.lan@m.scnu.edu.cn.*

*S1. Site-Exciton Models.*

In current work, we employed four site-exciton models (Model I - IV) and their parameters are given in Table S1. For all of them, we took the frequency domain as 0-1200 cm$^{-1}$ and Δω=12 cm$^{-1}$ to build a set of discreted bath modes.

Table S1. Different site-exciton models with their parameters.

|  | $V_{11} - V_{12}$ (eV) | $V_{12}$ (eV) | $\omega_c$ (cm$^{-1}$) | $\lambda$ (cm$^{-1}$) |
|---|---|---|---|---|
| Model I | 0 | 0.0124 | 200 | 64 |
| Model II | 0 | 0.0124 | 200 | 256 |
| Model III | 0.0186 | 0.0124 | 200 | 64 |
| Model IV | 0.0186 | 0.0124 | 200 | 225 |

*S2. LSTM-RNN Model Structures obtained by Hyperparameter Optimizations.*

Table S2. The neuron number of LSTM layers in the optimized LSTM-RNN model structures obtained from the hyperparameter optimization. For the dense layer, we used three neurons since three elements were predicted.

|  | SA | BO-TPE | RS |
|---|---|---|---|
| 2LSTM layers | 130-70, 210-110, 210-430, 310-270 | 370-210, 10-90, 70-110, 70-330 | 510-290, 370-190, 10-230, 110-110, 70-330, 490-390, 230-250 |
| 3LSTM layers | 350-50-430, 290-30-290, 110-390-450, 290-150-170, 130-50-190, 510-190-510, 410-290-350 | 70-130-350, 410-130-170, 330-390-450, 230-450-450, 190-50-130, 110-210-370 | 210-170-70, 90-230-290, 190-170-350, 430-250-170, 110-370-390, 350-310-70, 270-290-450 |
| 4LSTM layers | 50-170-470-270, 250-430-330-270, 90-250-290-250, 90-490-370-430, 90-290-150-270, 210-50-90-250, 30-250-150-330, 50-110-90-390, 330-310-30-410 | 310-90-410-390, 30-470-90-330, 490-210-270-30, 510-230-130-410, 130-130-110-90, 450-290-510-470, 110-70-130-390, 250-330-50-290, 170-30-270-450, 470-350-410-170 | 470-130-150-510, 170-450-450-330, 330-170-450-30, 510-230-90-230, 50-290-50-430, 150-290-410-190 |

## S3. The Convergence Tests of the LSTM-RNN Prediction with Bootstrap.

The main manuscript gives the results simulated by the (SA/BO/RS-H10)×BT100 LSTM-RNN models. Here two additional models are considered to check convergence, which are the (SA/BO/RS-H5)×BT100 and (SA/BO/RS-H1)×BT100 LSTM-RNN models. All results are given in Figure S1-S4. By comparison, the selection of 10 models from the hyperparameter optimization is enough to achieve convergence.

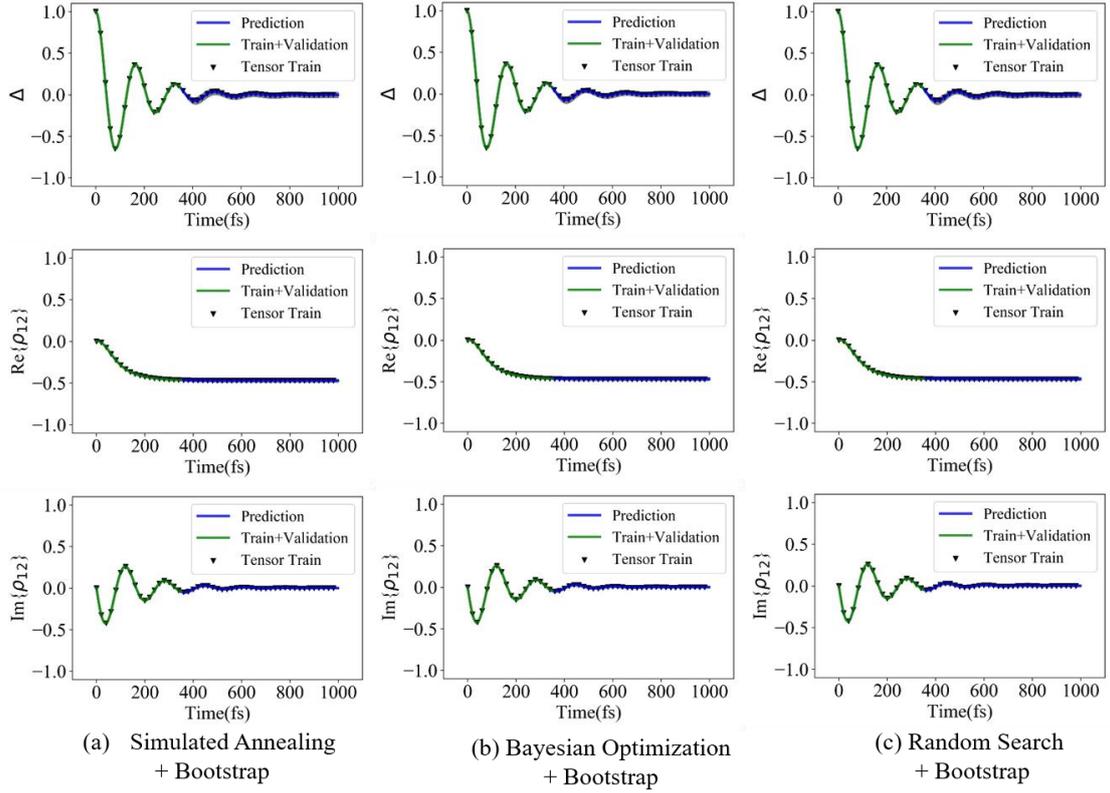

Figure S1. The quantum dynamics simulated by the (SA-H5)×BT100 (a), (BO-H5)×BT100 (b), and (RS-H5)×BT100 (c) LSTM-RNN models *vs.* the tensor-train quantum propagation in Model I. The green lines denote the training and validation samples (<350 fs) used in the LSTM-RNN model construction. The black triangles display the tensor-train simulation results. The blue lines correspond to the LSTM-RNN prediction of the future dynamics and the grey region shows the prediction uncertainty.

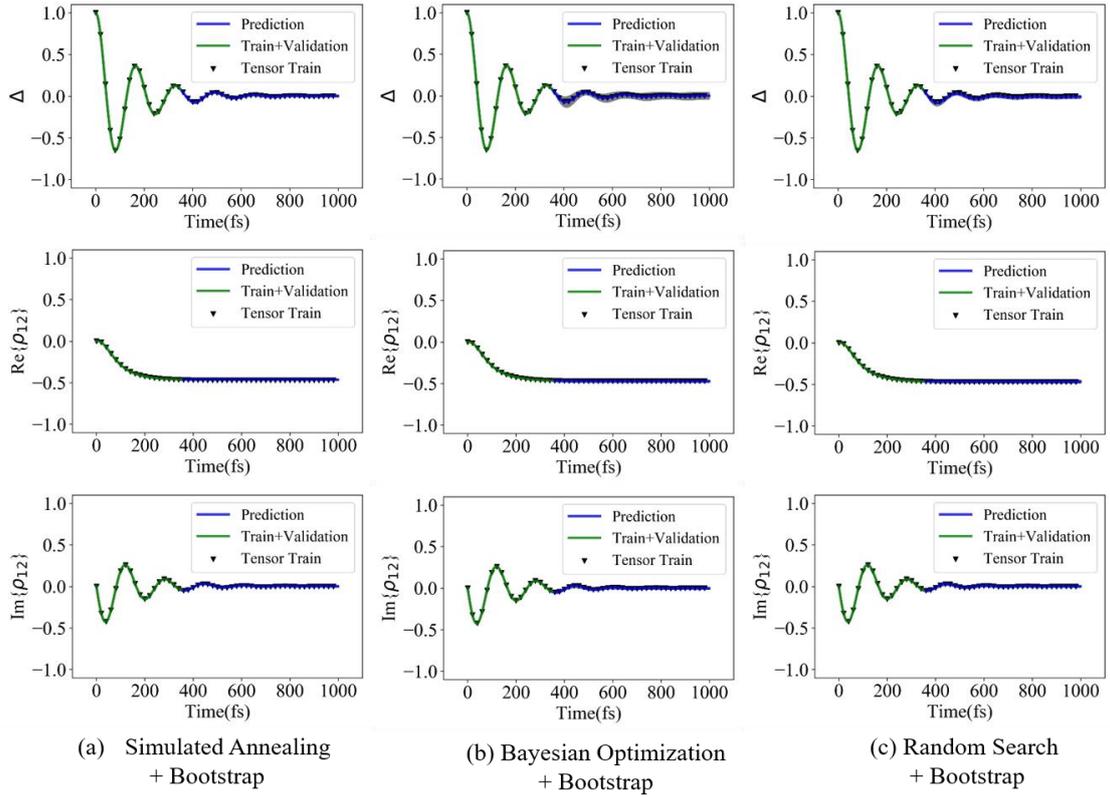

Figure S2. The quantum dynamics simulated by the (SA-H1)×BT100 (a), (BO-H1)×BT100 (b), and (RS-H1)×BT100 (c) LSTM-RNN models *vs.* the tensor-train quantum propagation in Model I. The green lines denote the training and validation samples (< 350 fs) used in the LSTM-RNN model construction. The black triangles display the tensor-train simulation results. The blue lines correspond to the LSTM-RNN prediction of the future dynamics and the grey region shows the prediction uncertainty.

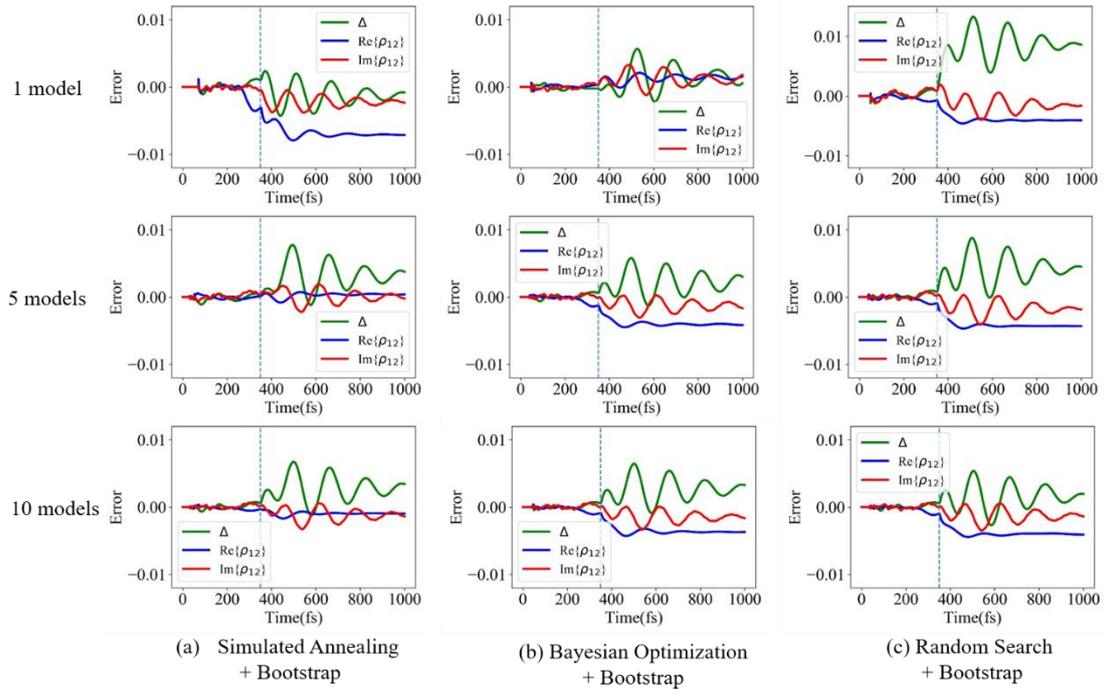

Figure S3. The prediction error in the different bootstrap-based LSTM-RNN simulation of the quantum dynamics including (SA-H1/H5/H10)×BT100 (a), (BO-H1/H5/H10)×BT100 (b), and (RS-H1/H5/H10)×BT100 (c). The blue dotted lines denote the training and validation samples (<350 fs) used in the LSTM-RNN construction.

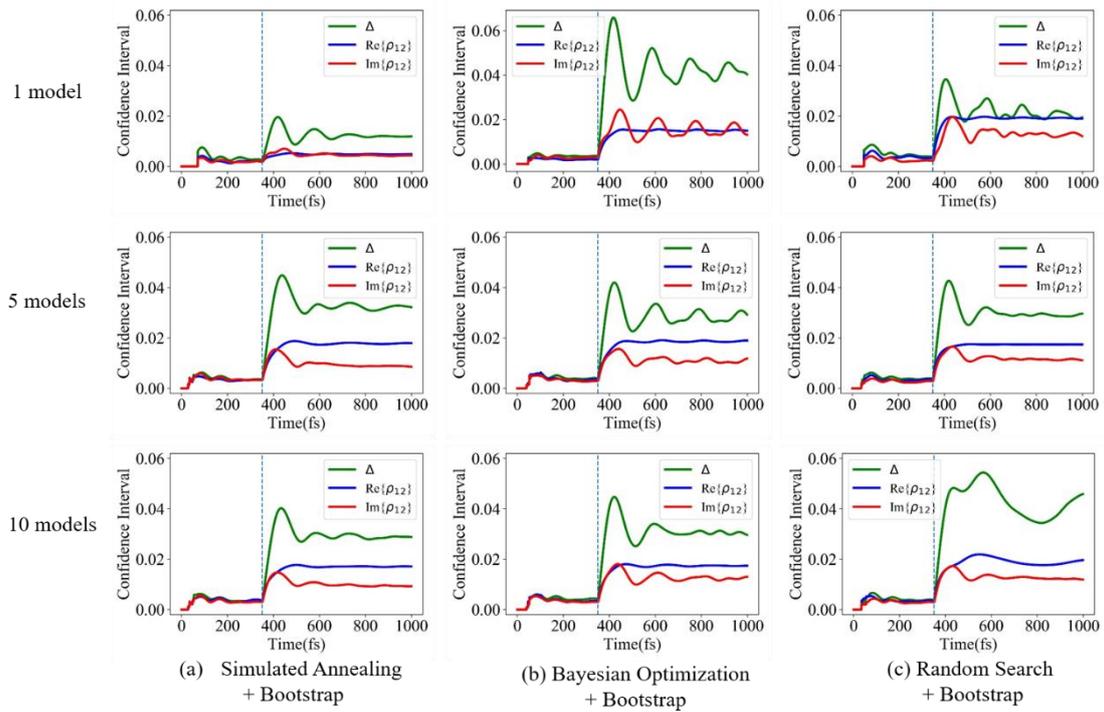

Figure S4. The confidence interval in the different bootstrap-based LSTM-RNN simulation of the quantum dynamics including (SA-H1/H5/H10)×BT100 (a), (BO-H1/H5/H10)×BT100 (b), and (RS-H1/H5/H10)×BT100 (c). The blue dotted lines denote the training and validation samples (<350 fs) used in the LSTM-RNN construction.

## S4. The RS + MC Dropout LSTM-NN Prediction.

When we combined the random search hyperparameter optimization and the MC dropout approach, both prediction error and uncertainty become very large, as shown in Figure S5.

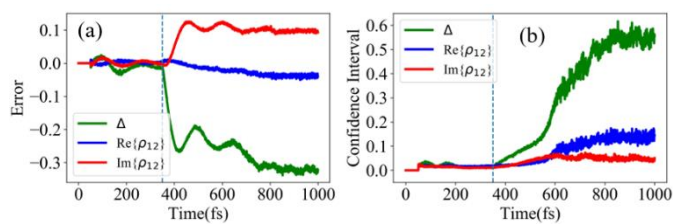

Figure S5. The prediction error (a) and confidence interval (b) of the (RS-H1)×MC100 LSTM-RNN simulation of the quantum dynamics of Model I. The blue dotted lines denote the training and validation samples (<350 fs) used in the LSTM-RNN construction.

## S5. The Convergence Tests of the LSTM-RNN Prediction by Combining Bootstrap and MC Dropout.

We combined the bootstrap resampling and MC dropout approaches to estimate the prediction confidence interval. In the main manuscript, the (SA/BO/RS-H1)×BT50 ×BT50 LSTM-RNN models were used, and totally 2500 networks were considered. Here we only select 1000 LSTM-RNN models randomly from them and conduct the prediction. All results are given in Figure S6 and Figure S7. Comparing two cases, the employment of 2500 networks should be enough to achieve the convergence.

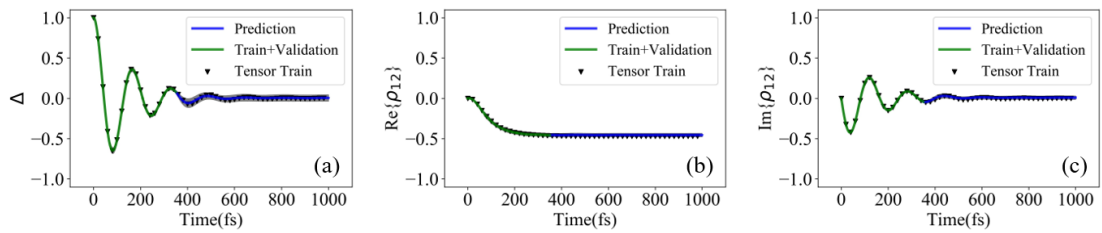

Figure S6. The quantum dynamics modelled by the (SA-H1)×BT10×MC10 LSTM-RNN models *vs.* the tensor-train quantum propagation in Model I. The green lines denote the training and validation samples (<350 fs) used in the LSTM-RNN construction. The black triangles display the tensor train simulation results. The blue lines correspond to the LSTM-RNN prediction of the future dynamics and the grey region shows the prediction uncertainty.

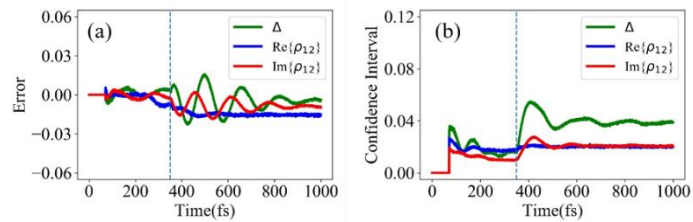

Figure S7. The prediction error (a) and confidence interval (b) of LSTM-RNN quantum dynamics propagation by using the combination of bootstrap and MC dropout with 1000 models. The blue dotted lines denote the training and validation samples (<350 fs) used in the LSTM-RNN construction.

## S6. The Prediction Error and Confidence Interval of the Other Models.

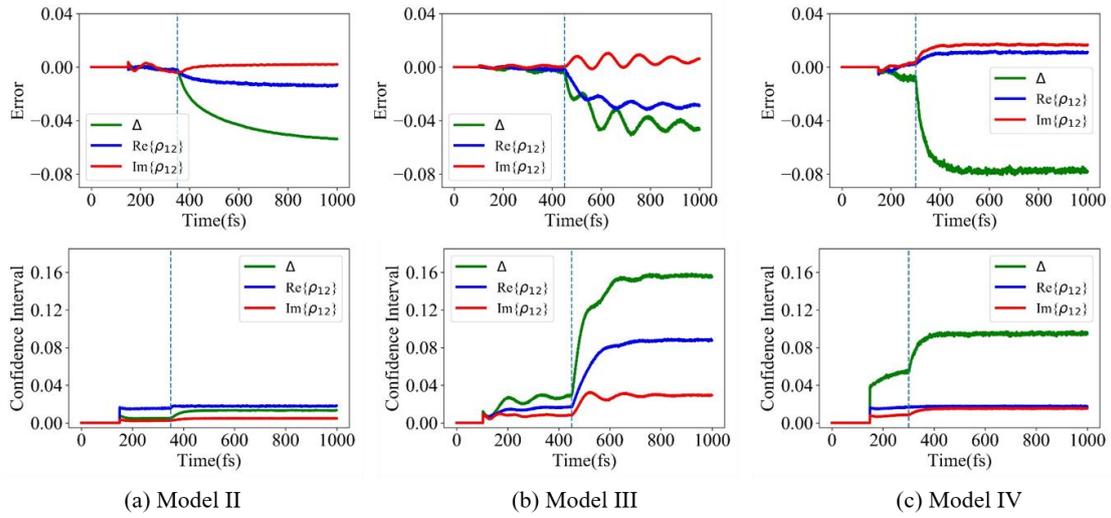

Figure S8. The prediction error and confidence interval in the (SA-H1)×BT50×MC50 LSTM-RNN simulation which combines with bootstrap and MC dropout of the quantum dynamics in (a) Model II; (b) Model III; (c) Model IV. The blue dotted lines denote the training and validation samples (<350 fs (a), <450 fs (b), <300 fs (c)) used in the LSTM-RNN construction.